\documentstyle[12pt]{article}

\begin{document}

\baselineskip=7mm

\begin{center}

{Many-body Systems Interacting via a 
Two-body Random Ensemble:
average energy of each angular momentum}

{ Y. M. Zhao$^{a,c}$,
  A. Arima$^{b}$, and N. Yoshinaga$^a$}

\vspace{0.2in}
{ $^a$ Department of Physics, Saitama University, Saitama-shi, Saitama 338 Japan \\
$^b$ The House of Councilors, 2-1-1 Nagatacho, 
Chiyodaku, Tokyo 100-8962, Japan \\
$^c$ Department of Physics,  Southeast University, Nanjing 210018 China }

\date{\today}

\end{center}

\begin{small}

In this paper, we discuss the  regularities of
 energy of each angular momentum $I$
 averaged over all the
states  for a fixed angular momentum (denoted as $\bar{E}_I$'s) 
  in many-body systems  interacting via a  
two-body random ensemble.
 It is found that  $\bar{E}_I$'s with
   $I \sim I_{min}$ (minimum of $I$) or $I_{max}$   have   large
probabilities (denoted as ${\cal P}(I)$) to be the lowest, and that 
${\cal P}(I)$ is close to zero elsewhere. 
A  simple argument based on the randomness of the two-particle  cfp's is given.
A compact trajectory  of the energy $\bar{E}_I$ vs. $I(I+1)$ 
is found to  be robust. 
Regular fluctuations of  the $P(I)$  (the probability
of finding $I$ to be the ground state)  and ${\cal P}(I)$ 
 of  even fermions   in a  
  single-$j$ shell and boson systems are found to be reverse,  and   
argued by the dimension fluctuation of the model space.
Other regularities, such as
why there are 2 or 3 sizable ${\cal P}(I)$'s with $I\sim I_{min}$ and
  ${\cal P}(I) \ll  {\cal P}(I_{max})$'s
   with $I\sim I_{max}$, why the coefficients $C$ defined by 
$\langle \bar{E}_I \rangle$=$CI(I+1)$ is sensitive to the orbits
and not sensitive to particle number, are found and studied
for the first time.  

\end{small}

        
\newpage

\section{Introduction}

The discovery  that 
 the  $0^+$  dominance in ground states (0 g.s.)
 \cite{Johnson1}  of even
fermion systems  can be obtained by
using  a  two-body random ensemble (TBRE) brought the physics of many-body
systems interacting via a TBRE, such
as the origin of this $0^+$ dominance and  its physical
implications into a sharp focus [2-18].
However, a sound understanding of this $0^+$ dominance is not
yet available, though
quite a few approaches were suggested.  
In this paper we discuss many-body
systems interacting via a TBRE hamiltonian in another
context: regularities of averaged energy
(denoted as $\bar{E}_I= \frac{1}{dim_I} \sum_{\beta} E_{I\beta}$,
where $\beta$ represents additional quantum numbers of angular momenta $I$
states and $dim_I$ is the dimension of $I$ states)
for  each  angular  momentum $I$.

Our study of $\bar{E}_I$   originates from
the difficulties \cite{prc1,Zhaox,Arima} that
we met in  studying  the $P(I)$, the probability
of finding angular momentum $I$ to be  the ground state. 
In Ref.  \cite{Arima} we studied 
the  $P(I)$'s in two steps. The first step  is  to study the
${\cal P}(I)$'s, the probabilities of $\bar{E}_I$'s
being the lowest. It is much simpler to understand
the behavior of  ${\cal P}(I)$'s 
than to understand the behavior of $P(I)$'s, because 
the  $\bar{E}_I$'s  are always  linear combinations
of two-body matrix elements,  and  the approach of \cite{Zhaox} 
is applicable to  predict the  ${\cal P}(I)$'s of
a randomly interacting many-body system.
The second step is to  study 
 the  features of width for each $I$. 
Unfortunately, such a picture is not applicabe to
odd number of fermions.
The reason is that the correlation between states 
is also essential to explain the distribution of $P(I)$'s.
Here correlation is an antonym  of independence, it 
refers to, e.g.,  for fermions in a single-$j$ shell,
the state with $I_{max}-2$ is  very likely the first
excited state when the $I=I_{max}$ state is the g.s..
It is not enough  to consider only 
the statistical behavior of energy levels, especially
for systems with few parameters (e.g., $sd$ boson systems, fermions
in a single-$j$ shell).    

Although we did not successfully relate the ${\cal P}(I)$'s
with  $P(I)$'s in \cite{Arima}, regularities of
the ${\cal P}(I)$'s  studied in \cite{Arima} 
are very interesting and worthy of further study, as will be shown
in this paper.

This paper is organized as follows: In Sec. 2, we present
a few typical examples of  ${\cal P}(I)$'s, which
show that the
  $\bar{E}_I$'s with
   $I \sim I_{min}$ (minimum of $I$) or $I \sim I_{max}$   have   large
probabilities (denoted as ${\cal P}(I)$) to be the lowest, and that 
${\cal P}(I)$ is close to zero elsewhere. A plausible understanding
is given in terms of distributions of two-body coefficients of
fractional parentage. In Sec. 3, we discuss  
a robust  correlation between $\bar{E}_I$'s and $I(I+1)$ for all cases
that we checked:
 fermions in  a single-$j$ 
or a two-$j$ shell, $d$-, $sd$-, and $sdg$-boson systems. 
Furthermore, we define a ``moment of inertia", ${\cal J}$, by
${\cal J}=$ $\frac{1}{2} I(I+1)/  \langle  \bar{E}_I \rangle$, and find that
$\sqrt{\cal J}$ is proportional to $j$ of fermions in a single-$j$ shell
(and $d$- and $sd$-boson systems), 
and almost independent of particle number $n$. 
Similar behavior is noticed  in $sdg$-boson systems and fermion systems
in two-$j$ shell. 
Here, we find a regular even-odd staggering of  both ${\cal P}(I)$'s
and $P(I)$'s  of even fermions in a single-$j$ shell, and
 $d$-, $sd$-, and $sdg$-boson systems, 
which is  argued by a regular odd-even staggering
of  dimension of the model space.
One apparent observation that 
 there are 2 or 3 sizable ${\cal P}(I)$'s with $I\sim I_{min}$ and
 only one large ${\cal P}(I_{max})$  with 
  ${\cal P}(I)s \ll  {\cal P}(I_{max})$ $(I \sim I_{max})$
  is interpreted as a reflection of
   fluctuations of $\bar{E}_I$ in each run of a 
   TBRE hamiltonian. 
In Sec. 4 we reproduce  the relation of $\bar{E}_I$'s and $I(I+1)$ and
explain the features of ${\cal J}$'s for fermions in a single-$j$ shell
by assuming randomness of two-body cfps. 
A summary of this work will be given in Sec. 5.
                                               
\newpage

\section{Typical examples of ${\cal P}(I)$'s}

In Fig. 1  we plot  the ${\cal P}(I)$'s for a few very different
systems: 4 fermions in a single-$j$ shell ($j=\frac{15}{2})$,
 6 fermions   in a two-$j$  ($2j_1, 2j_2$)=(11,7) shell, 
    a 6-$sdg$-boson system, and 5 fermions in
  a single-$j$ shell ($j=\frac{9}{2})$. The two-body interaction parameters 
are taken as the two-body random ensemble (TBRE), as
most of  previous works [2-18]. All the  ${\cal P}(I)$'s are
obtained by 1000 runs. 
We show in Fig. 1  typical examples among
many cases that we have  checked:
  4, 5, and 6 valence fermions in a single-$j$ shell 
    up to $j=\frac{31}{2}$,  
both  even or odd number of  fermions ($n=4$ to 9) in
   two-$j$ shells with  ($2j_1, 2j_2)$=(7,5), (11,3), (11, 5), (11,7), (11,9),
(13,9),  $d$-boson systems with $n_d$ changing from 3 to
45, $sd$-boson systems with $n$ changing from
4 to 17, and $sdg$-boson systems with
$n=4, 5, 6$. It is noted that  
the ${\cal P}(I)$'s are   large
if $I \sim I_{min}$ or $I_{max}$; they are
close to zero elsewhere. This feature is
general for all cases. It is noted that
a displacement on the TBRE produce only a constant shift on the
ground state energy. Thus it is
robust regardless of the displacement, while
$P(I)$ is quite sensitive to the displacement \cite{Zuker}. 

Now  we give an understanding on this observation.  
 Let   $\bar{E}_I = \sum_{k_0} \bar{\alpha}_I^{k_0} G_{k_0}$,
 where $G_{k_0}$'s are
two-body matrix elements  determined randomly (TBRE parameters),
and $\bar{\alpha}_I^{k_0}$'s are coefficients calculated in terms of   
two-particle cfp's.
For the sake of simplicity, here we restrict ourselves to fermions in a
single-$j$ shell 
but a generalization to other cases, we believe, is straightforward.
In a single-$j$ shell,  $k_0$ is uniquely labeled by  
  $J$ \cite{Zhaox}. We use    $G_J$ and $\bar{\alpha}^J_I$ instead of
$G_{k_0}$ and $\bar{\alpha}^{k_0}_I$   for a single-$j$ shell. 
As a specific example, we
discuss the case of $j=\frac{9}{2}$ with 4 fermions below.

Using the  coefficients $\bar{\alpha}^J_I$,
we   predict  the ${\cal P}(I)$'s by using  integrals similar to
Eq. (7) of Ref. \cite{Zhaox}, without diagonalizing 
a TBRE hamiltonian. 
The predicted ${\cal P}(I)$'s for 4 fermions in a  $j=\frac{9}{2}$ shell 
 are listed in the column ``pred1." in Table I. It is easy to
notice  that  the  ${\cal P}(I)$ is large 
 if $\bar{E}_I$  has one or more
largest (and the smallest) $\bar{\alpha}^J_I$'s  for different $I$'s 
but a fixed $G_J$. The predicted ${\cal P}(I)$'s 
by the simple approach proposed in \cite{prc1}, 
${\cal P}(I)=$${\cal N}_I/N_m$, 
  are   listed in the
column ``pred2." of  Table I, where
$N_m=2N-1$, and ${\cal N}_I$ is the number of angular momenta $I$
being either the smallest or the largest eigenvalues with
only one of two-body matrix elements switched on. 
The ${\cal P}(I)$'s obtained
by diagonalizing  a TBRE hamiltonian (1000 runs) are given
in the column ``TBRE".  It is seen that
all predicted ${\cal P}(I)$'s are reasonably consistent with
those obtained by diagonalizing a TBRE hamiltonian.

It is interesting to  ask 
 why  the largest $\bar{\alpha}^J_I$'s
(for different $I$ and fixed $J$) appear in $\bar{E}_I$'s 
with $I\sim I_{min}$ or $I_{max}$.  Our interpretation is as follows.

First,    $\bar{\alpha}^J_I$'s
 are  the largest if  $I\sim I_{min}$ (small $J$ cases)  or $I_{max}$
 (large $J$ cases) 
by assuming that  the two-particle cfp's, and hence 
the  ${\alpha}^J_{I \beta \beta'}$'s
 are {\bf randomly} given. This assumption  was found  \cite{Arima}
to   work  well except at the edge of angular momenta $I$.
As a specific example, we show  the distribution of cfp's and
$\alpha_{I \beta \beta'}$'s for 4 fermions
in a $j=\frac{31}{2}$ shell in Fig.  2.
It is interesting to note  that
both the cfp's and ${\alpha}^J_{I \beta \beta'}$'s
seem to follow a Poissonian distribution, the origin
of which is not known.

We thus define $d_I^J/D_I$, a ratio between the dimension of 
  two-particle cfp's corresponding to  
interaction $G_J$ of angular momentum $I$ state  and
the total dimension of the two-particle cfp's of angular momentum $I$ state.
The  $d_I^J/D_I \times \frac{1}{2} n (n-1)$ is then a reasonable estimate
of $\bar{\alpha}^J_I$ \cite{Zhaox,Lawson} if one 
assumes  the two-particle cfp's are ``uniformly" distributed
in the $D_I$ components.
It is easy to realize  that 
 $\bar{\alpha}^J_I$'s  for states with  
small $J$ and  $I\sim I_{min}$ are larger 
 than those of  other states because   
 $d_I^J/D_I$ of small $J$ and  $I\sim I_{min}$  states are
 mostly larger. Similar statements  are applicable to the
$\bar{\alpha}^J_I$'s of the states with
$I\sim I_{max}$ and large $J$.
To exemplify this argument, in Table II we list  $d_I^J$ and $D_I$
of 5 fermions in a  $j=\frac{15}{2}$ shell, where the $d_I^J/D_I$
with $J < j-\frac{1}{2}$ and $I \sim I_{min}$ 
 are mostly larger  than that of states elsewhere,
 and   the $d_I^J/D_I$ with $J > j-\frac{1}{2}$
 and $I \sim I_{max}$ are larger than that of other states.

Second, the fluctuations of $\bar{\alpha}_I^J$ with
$I\sim I_{min}$ or $I_{max}$ are larger than those with medium $I$. 
In all cases, including boson systems,
 even or odd number of fermions in a  single-$j$ or a many-$j$ shell,
the dimension of two-particle cfp's and that of the
model space with $I \sim I_{min}$ or $I_{max}$ 
are smaller than those with medium $I$, which suggests that  
 the fluctuations of
the $\bar{\alpha}^J_I$ with $I \sim I_{min}$ or
$I_{max}$ are much larger than those elsewhere if it is assumed  
that the two-particle cfp's are ``random".

  Therefore, the statistical properties  of two-particle cfp's 
and  the dimension   of two-particle cfp's,  and the property 
  of the shell model dimension accounts
  for the facts why there are always many $\bar{\alpha}^J_I$'s,
  which are the largest (or the smallest) for different $I$ and fixed $J$, 
in  $\bar{E}_I$'s with $I \sim I_{min}$ or $I_{max}$.

A  ``staggering" of ${\cal P}(I)$'s
is noticed for boson systems and
even fermion in a single-$j$ shell:
the ${\cal P}(I)$'s of
odd $I$ are  larger than their even neighbors
with very few exceptions. 
In contrast to this systematics,
it was noticed \cite{zelevinsky,Zhaox} that the ${P}(I)$'s  for even $I$
in these systems are  larger than their odd neighbors. 
These two ``reversed" regularities
are explained as a reflection of a regular change
in dimension of the model space below: 
For boson systems and even fermions in a single-$j$ shell,
the dimension of the model space for even $I$ is  systematically and
relatively larger
than their odd $I$ neighbors.  The ${\cal P}(I)$'s are determined by  
the coefficients $\bar{\alpha}^k_I$.  The relatively larger  dimensions of the
model space for even $I$'s equivalently lead to
relatively smaller fluctuations of $\bar{\alpha}^k_I$ of even $I$'s,
and the $\bar{\alpha}^k_I$ are relatively ``medium" compared with their
odd-$I$ neighbors. This gives relatively larger ${\cal P}(I)$'s for
odd $I$'s. On the other side, the  ${P}(I)$'s are determined
 by the largest eigenvalues with one of the two-body
  matrix elements switched on. The 
  larger dimension of the
  model space for even $I$ states equivalently
produces a relatively larger maximum (denoted as
$E_I^0$) of  $E_I$'s  for even $I$'s than
their odd neighbors, where the $E_I^0$'s are obtained by
diagonalizing two-body matrix elements $N$ times, in  each time only
one of $G_k=1$ and the other interactions are switched off. 

\newpage

\section{A regularity of   averaged energies of fixed
angular momenta}

If one examines the spin ordering of the  average 
energies $\bar{E}_I$'s, it is easy to notice  that if 
 the angular momentum $I$ of the lowest $\bar{E}_I$ belongs to the
 case of $I\sim I_{min}$ ($I_{max}$) in {\bf one} run
 of the TBRE,  very likely the average energy $\bar{E}_I$
 increases  (decreases) 
 with the total angular momentum $I$, and is proportional
 to $I(I+1)$ on average. The probability
 of $\bar{E}_I$ being the lowest with ``medium" $I$ is close to zero.

Let $ \langle \bar{E}_I\rangle_{min}$ 
($\langle \bar{E}_I\rangle_{max}$) be a quantity obtained by  
averaging $\bar{E}_I$  over the cases
with only $I\sim I_{min}$ ($I\sim I_{max}$)  g.s. 
for the ensemble used.  We find that both
$\langle \bar{E}_I\rangle_{max}$ and $\langle \bar{E}_I\rangle_{min}$
are proportional to  the  $I(I+1)$,   similar to
a ``rotational" spectra. 

In Fig.  3 we  show  $\langle \bar{E}_I\rangle_{min}$ vs. $I(I+1)$ for
a  $d$-boson system with $n=20$, a  $sd$-boson  system with $n=10$, and
 a   system with 4 fermions in a single-$j$ ($j$=17/2) shell, 
 and a  system with 6 fermions in a two-$j$ shell.
For the sake of simplicity, below we introduce ``moment of
inertia" ${\cal J}$, defined by a 
optimal constant $(2C)^{-1}$, where $C$ is the
coefficient defined in $\langle \bar{E}_I\rangle_{max}$
($\langle \bar{E}_I\rangle_{min}$) $=C I (I+1)$.

In the pioneering paper \cite{Johnson1} by Johnson et al,
they showed in their Fig. 4 that the average excitation energy
follows roughly a function proportional to
$I(I+1)$, but the energy they refered to are
average energy of yrast and even $I$ states, therefore it is
different from the $\langle \bar{E}_I\rangle_{min}$  in
this paper. It is noted that we consider
both even $I$ and odd $I$ in our ``rotation". 

Prior to this work, based on geometric chaoticity, 
 Mulhall, Zelevinsky, and
Volya \cite{zelevinsky}
elaborated a formula below:
\begin{equation}
   E_I  = {\rm Const}_1 + {\rm Const}_2 I(I+1)
  + o(I(I+1)).
\end{equation}
Those  authors discussed the
 $P(I)$'s by using the above formula for fermions
 in a single-$j$ shell. 
It seems, however, that the $E_I$ in their formula 
should correspond to our $\langle E_I \rangle$. 
The features  predicted by \cite{zelevinsky}, such
 that the minimum  and the maximum of  
angular momenta $I$ are favored to be the ground states with 
around 50$\%$ for each
(sum of all  ${\cal P}(I)$'s with $I \sim I_{min}$ or
$I_{max}$), and that the $\langle \bar{E}_I\rangle_{min}$'s  (and $\langle \bar{E}_I\rangle_{max}$)
behave like a  rotor, seem more appropriate to  
describe  the ${\cal P}(I)$'s,
$\langle \bar{E}_I\rangle_{min}$ and $\langle \bar{E}_I\rangle_{max}$, rather than the
 $P(I)$'s and the energy obtained by  averaging yrast energy  over the cases
 of 0 g.s. or $I_{max}$ g.s. in the ensemble.
The ``equilibrium" energies obtained
in the statistical way of \cite{zelevinsky}
may correspond to the energies averaged  over
all states with given angular momenta $I$'s, not the yrast ones.
In fact, the  $I(I+1)$ behavior  in 
  average yrast energy defined in ref. \cite{zelevinsky}
was not well confirmed. 

In another recent work \cite{Zuker}, the spectroscopy
of $\bar{E}_I$ was  checked but this  
pattern was not noticed.

It is noted that one should not
confuse  $\bar{E}_I$'s
 with Bethe expression of level densities \cite{Bohr} which is 
 is based on a Fermion gas approach. 
First, the  $\bar{E}_I$'s are always Gaussian
 and  $\bar{E}_I$'s obtained by averaging  over a TBRE  
should be  zero.  The  $I(I+1)$  behavior appears when
one divides a TBRE into two cases--$I\sim I_{min}$ g.s. and
$I\sim I_{min}$ g.s.-- and   calculates 
$\langle \bar{E}_I\rangle_{min}$ and $\langle \bar{E}_I\rangle_{max}$. 
Second, it is emphasized that an $I(I+1)$
behavior of $\bar{E}_I$ in  the Bethe expression of
level density   and  that in $\langle \bar{E}_I \rangle_{min}$
(or $\langle \bar{E}_I \rangle_{max}$) 
 are   completely  different.
For example,   ${\cal J}$ in Bethe expression changes
with particle number but ${\cal J}$ in this work, as will be shown,
is not sensitive to particle number $n$.
Furthermore,   systems discussed in this paper can be very simple, and
those described by Bethe expression require complexity
in energy levels so that one
needs statistical approach for level densities. 
The $I(I+1)$ behavior discussed in this paper was  not reported in 
previous works. 
             
In Fig. 4 we plot $C$  for 
  fermions in a few single-$j$ or two-$j$ shells.
  The results  indicate: 

1. ~ $C$ is almost independent of particle number $n$ for all systems. 

2. ~$C$ of fermions in a single-$j$ shell 
 is quite sensitive to orbits (labeled by $j$),
 that of fermions in a two-$j$ shell is sensitive to 
the larger $j$ value.  
The coefficient $C$ decreases with the  larger   $j$ 
if the smaller $j$'s are the same. For instance,  coefficient 
$C$ of ($2j_1, 2j_2)=(7,5)$, that of ($2j_1, 2j_2)=(9,5)$
and that of ($2j_1, 2j_2)=(11,5)$ decrease successively. 
                     
3. ~For fermions in a two-$j$ shell,
the difference between coefficients $C$
of systems with the same  $j$ (larger) and 
$j_2 \ll j_1$ is very small. For example, the coefficients $C$
for shells  ($2j_1, 2j_2)=$ (11,1), (11,3)  are very close.

 Without presenting details we mention that the
$C$ determined by
$\langle \bar{E}_I\rangle_{min}$ $=C I (I+1)$ is  close to
that by  $\langle \bar{E}_I\rangle_{max}$  $=C I (I+1)$
 for all examples that we have checked.  This
could be easily understood by the symmetry of the ensemble that
we use.

 An empirical relation between  ${\cal J}$ and $j$ is
summarized in Fig. 5. For fermions in a single-$j$ shell, 
  $d$ boson systems and $sd$ boson  systems, ${\cal J}$
  is fitted by a trajectory of $\sqrt{{\cal J}}= 1.42 j$
  (we take $j=2$ for $d$ boson and $sd$ boson
systems); for $sdg$ bosons and fermions in a two-$j$ shell, there seems
a slight shift from the trajectory  $\sqrt{{\cal J}}= 1.42 j$,
where we take $j^2= j_1^2 + j_2^2$.  It is noted that one cannot, though
the  Eq. (7) of  \cite{zelevinsky} seems to indicate  a correct behavior
of $\langle \bar{E}_I\rangle$, obtain this simple systematics of 
${\cal J}$ based on their formulas.
The ${\cal J}$ of \cite{zelevinsky}   is very different from
that given here. Therefore, this regularity
cannot be  explained by previous schemes such as
\cite{zelevinsky}  and deserves further
studies. 

Another feature of ${\cal P}(I)$'s is that 
   that the ${\cal P}(I_{max})$'s are always 
quite ``stable"   ($\sim$28-35$\%$), while 
 the ${\cal P}(I_{max}-2)$'s of fermions  in a 
  single-$j$ shell   and boson systems, and
  ${\cal P}(I_{max}-1)$'s of fermions in a many-$j$ shell, 
    is drastically smaller 
   than those of the  $I_{max}$ state(s), though  still
   sizable     ($\sim 5-15 \%$), and that 
  there may be 2 or 3 sizable
${\cal P}(I)$'s for the cases of $I\sim I_{min}$, and 
the ${\cal P}(I_{min})$'s are not {\it always}   larger than other
${\cal P}(I)$'s  (with $I \sim I_{min}$).

Now  we are able to explain  the asymmetry of  ${\cal P}(I)$ by 
using the fluctuations of $\bar{E}_I$. 
The $\bar{E}_I$'s are   proportional to 
$I(I+1)$ but with fluctuations in each run  of a TBRE hamiltonian.
Because   $ \left( \bar{E}_{I+1} - \bar{E}_I \right)$
is small if $I$ is small  and large if $I$ is large,
the probabilities to change order of
 $\bar{E}_I$'s for   $I \sim I_{min}$ due to the fluctuations of $\bar{E}_I$ 
 is much larger  than 
those for $I \sim I_{max}$. That is why there are 2 or 3 sizable
and comparable ${\cal P}(I)$'s with $I\sim I_{min}$ but
only one large 
${\cal P}(I)$'s with $I\sim I_{max}$ and the  
${\cal P}(I_{max}-1) \ll {\cal P}(I_{max})$.

\newpage

\section{A scenario  of 
  $\langle \bar{E}_I \rangle \sim I(I+1)$ relation}

It is more convincing if one ``reproduces" the 
 $\langle \bar{E}_I \rangle I(I+1)$ relation by the above
 assumption of randomness of two-body cfp's. 
Below we first
show by one specific case that
   $\langle \bar{E}_I \rangle$ vs.  $I(I+1)$
for  fermions in a single-$j$ shell,
then we present a simple approach  to 
evaluate   ${\cal J}$ of fermions in a  single-$j$ shell.
Note that a simple approach to evaluate the  ${\cal J}$
of fermions in a many-$j$ shell  is not yet available.

As indicated in Fig. 4,  ${\cal J}$ is not sensitive to particle number
 $n$. Thus we study only 4-fermion systems. We concentrate on
states with $I \ge 2j-1$, where the
dimension of two-particle cfp's is very simple: 
\begin{eqnarray}
&& D_I = {\cal L} ({\cal L} +1)/2, \nonumber \\
&& d_I^J =
\left\{
\begin{array}{ll}
{\cal L}+ J/2 + (1 - 2j)/2, &  {\rm if}~ 2j \le {\cal L}+ J + 1 - 2j    \\
0  &  {\rm otherwise}
\end{array}  \right.;
\end{eqnarray}
where ${\cal L} = 2j - \left[ I/2 \right]$.  
Suppose that the two-particle cfp's are uniformly distributed, and that 
the number of two-particle cfp's are large enough for a statistical prediction.
We have $\bar{\alpha}_I^J$=$d^J_I/D_I \times n(n-1)/2$.
It was checked in \cite{Arima} that the $\bar{\alpha}_I^J$'s
predicted in this way is well consistent with those 
 by diagonalizing matrices if the dimension
 of two-particle cfp's is quite large.

Next we use these predicted   $\bar{\alpha}_I^J$'s and random
two-body interaction parameters (TBRE), $G_J$'s, to evaluate the 
averaged energies of given angular momenta.
As an example we use $j=15/2$ and $n=4$.
The results are shown in Fig. 6a). It is seen that the
averaged energies, $\langle \bar{E}_I \rangle$'s,
are approximately proportional to $I(I+1)$. 

Now let us introduce an approximation  of $\langle G_J \rangle_{min}$'s by
only 3 requirements for 4 fermions in a single-$j$ shell.
The first requirement is that  $\sum_j \langle G_J \rangle_{min} \sim 0$,
and the second requirement is that  $\langle G_{J_{max}} \rangle_{min}=0.7$,
$\langle G_{J_{max}-k} \rangle_{min} \sim 0.3$ with $k=2,4$.
The last requirement is  $\langle G_{J} \rangle_{min} \sim -0.3$
when $J < 8$. Other $\langle G_{J} \rangle_{min}$ is quite small
in magnitude. Those requirements are based on our calculations
of fermions in a single-$j$ shell. 
By using these  $\langle G_{J} \rangle$'s and
$\bar{\alpha}_I^J$'s, we obtain 
Fig.  6b). It is seen that the results are very similar:
The ${\cal J}$ obtained in Fig. 6a)
is 9.96, that obtained in Fig. 6b) is 11.18, and that
in Fig.  5, which is obtained by 
by diagonalizing a TBRE hamiltonian, 
 is 10.30. Therefore, we show clearly  by this example  that
 the $I(I+1)$ behavior of $\bar{E}_I$'s, 
  at least for 4 fermions in a single-$j$ shell, 
   is   a reflection of 
a random distribution of the two-particle cfp's.

Below we estimate    ${\cal J}$,
and thereby giving a simple argument on the
behavior of  ${\cal J}$ of fermions  in a single-$j$ shell.
We take that $\langle \bar{E}_{I_{min}} \rangle \sim 0$,
and $I_{min}(I_{min}+1) \sim 0$. We assume that
$\langle \bar{E}_{I_{max}} \rangle \sim \alpha^{J_{max}}_{I_{max}}
\langle G_{J_{max}} \rangle$. For the sake of simplicity,
we take that $\langle G_{J_{max}} \rangle=0.7$ for $n=4$ in all
 single-$j$ shells. Then we have that for $n=4$: 
\begin{equation}
\alpha^{J_{max}}_{I_{max}} \langle G_{J_{max}} \rangle \sim
\frac{1}{2 \cal J} (4j-6)(4j-5) \sim \frac{8}{ \cal J}   j^2. \label{estimate}
\end{equation}
From \cite{prc1}, it is seen that the
 $\alpha^{J_{max}}_{I_{max}}$ of $n=4$ saturates quickly at $\frac{29}{8}$. 
Then the left hand side $\sim \frac{29}{8} \times 0.7$ $\sim  2.54$.
We find that $\sqrt{\cal J}  \sim 1.77 j$, a bit larger than
(but close to) the $\sqrt{\cal J}$  obtained in Fig. 5  
($\sqrt{{\cal J}}= 1.42 j$).   
 The   ${\cal J}$
in Eq.~(\ref{estimate}) is over-estimated
because we neglect the contributions from
$G_{J_{max}-2}$  and $G_{J_{max}-4}$ in this simple phenomenology.

The  ${\cal J}$ of $d$- and $sd$-boson systems may be
estimated in a similar way.   For example, $c_4 - c_2 
>0$ gives $\bar{E}_I$ with $I \sim 0$ ground states.
The  $c_4 - c_2  >0$ has an average around 0.88.
From Eqs.~(10-11), one has that ${\cal J} \sim 7/(c_4 - c_2)
\sim 7.95$, or $\sqrt{ \cal J} \sim 2.82$, consistent with
that in Fig. 5 (2.67).

\newpage

\section{ Summary and discussion}

To summarize,   we have presented in this paper for the first time 
two main  robust regularities 
of many-body systems interacting via  a  two-body random ensemble:
1. The   ${\cal P}(I)$'s,     probabilities of
average  energies $\bar{E}_I$ of the angular momentum $I$ states
being the lowest  for many-body systems,  are
large only  and only if  $I \sim I_{min}$ or $I_{max}$.
2. The  $I(I+1)$  behavior of
$\langle \bar{E}_I\rangle_{min}$ and $\langle \bar{E}_I\rangle_{max}$.
Without detailed discussions it is  noted 
in this paper that   the   regularities
of   ${\cal P}(I)$'s remain essentially
the same if one takes other ensembles, such as the displaced
random numbers, or random numbers which are only
positive (or negative), while those of 
$P(I)$'s by using other ensembles
 may be completely different from those obtained by a 
TBRE hamiltonian.  This suggests that  the regularities of
the ${\cal P}(I)$'s and  $\langle \bar{E}_I\rangle$ are very
{\bf robust}.

We first propose an approach  to interprete the  regularity
of ${\cal P}(I)$'s 
in terms of randomness of 
coefficients $\bar{\alpha}_I^k$, which are obtained
from the two-particle cfp's  and  the  approach developed in
Ref. \cite{Zhaox}. It is suggested that  the  dimension of two-particle 
cfp's and the that  of the  model space accounts for
the large $\bar{\alpha}_I^k$
with  $I \sim I_{max}$ and $I_{min}$.

The  ``staggering" patterns of ${\cal P}(I)$'s
and $P(I)$'s are found to be reverse 
 in boson systems and
even fermions in a  single-$j$ shell:
$P(I)$'s (${\cal P}(I)$'s) with even $I$
is systematically larger (smaller) than those of their odd $I$ neighbors. 
These staggering patterns are interpreted 
 in terms of a regular staggering in   dimension of the model space. 
 The asymmetry of  distribution  of  ${\cal P}(I)$'s
is explained by fluctuations of  $\bar{E}_I$.

We provide a scenario of ``reproducing" a compact trajectory of 
  $\langle \bar{E}_I\rangle$ plotted against  $I(I+1)$ by assuming the
randomness of two-particle cfp's in the case
of fermions in a single-$j$ shell.  We also propose a simple method to
estimate  ${\cal J}$  and a simple method to simulate 
  the linear relation between  $\sqrt{\cal J}$ of fermions in a
  single-$j$ shell  and the angular momentum of the orbit,  $j$.

 We therefore believe that the randomness of two-particle cfp's is the
origin of all the observed regularities related to
$\bar{E}_I$'s and ${\cal P}(I)$'s. 
                                    
Finally, it is pointed out 
that the ${\cal P}(I)$'s and $P(I)$ discussed
in our previous papers \cite{Zhaox,Arima} 
are  different quantities. For even systems
the behavior of these two  are accidentally    
similar. For odd-$A$ systems, however, the ${\cal P}(I)$'s
 are very different from  ${P}(I)$'s,
 which explicitly  demonstrates that
   the $I$ g.s. probabilities (and 0 g.s. dominance)  cannot be explained by 
 geometric  chaoticity \cite{zelevinsky}.  The 0 g.s. dominance is actually 
 related to two-body matrix elements which have specific features \cite{prc1}.

{\bf Acknowledgement:} 
One of us (YMZ) is  grateful to Drs. S. Pittel, 
 Y. Gono,   Y. R. Shimizu, and R. Bijker 
for discussions and/or communications.
This work is  supported in part by the Japan Society
for the Promotion of Science
under contract No. P01021.

\newpage

{TABLE  I. The coefficients $\bar{\alpha}_{I}^J$ 
for 4 fermions in a $j=\frac{9}{2}$ shell. 
 Bold font is used   for the largest
 $\bar{\alpha}^J_I$ are the largest and italic for
 the smallest  $\bar{\alpha}^J_I$ for a given $J$.  The probabilities
 in the column ``pred1." are obtained by  
 integrals similar to Eq. (7) in Ref. [17], and those
 in the column ``pred2." are obtained by
 the approach discussed in Ref. [16].
 The ${\cal P}(I)$'s in the last column ``TBRE" (in $\%$) 
  are  obtained by diagonalizing
 a TBRE hamiltonian 1000 runs. We take
 both the smallest and the largest $\bar{\alpha}^J_I$ 
 when counting ${\cal N}_I$. }

\begin{tabular}{ccccccccc} \hline  \hline
$I$   &  $G_0$ &  $G_2$ &  $G_4$ &  $G_6$ &  $G_8$ & Pred1.($\%$) & pred2.($\%$) & ${\cal P}(I)$  \\  \hline 
0 &  {\bf 0.80} & 0.35 & 1.74 & 2.11 & 1.01 & 11.97 & 11.1 & 10.2 \\
2 &  0.30 & {\bf 1.39} & 1.45 & {\it 1.29} & 1.56 & 14.51 & 22.2  & 15.4  \\
3 &  0.00 & 0.36 & {\bf 2.28} & {\bf 2.63} & {\it 0.71} &28.17 & 33.3 & 28.9 \\ 
4 &  0.20 & 1.07 & 1.38 & 1.91 & 1.44 & 1.74 &  0 &  1.7 \\
5 &  0.00 & 1.00 & 1.59 & 1.84 & 1.57 & 0.30 &  0 &  0.6 \\ 
6 &  0.20 & 0.79 & 1.50 & 1.58 & 1.93 & 0.22 &  0 &  0.3 \\  
7 &  0.00 & 1.20 & 1.09 & 1.40 & 2.31 & 3.44 &  0 &  3.2 \\  
8 &  0.30 & 0.48 & 1.05 & 1.82 & 2.36 & 0.03 &  0 &  0 \\  
9 &  0.00 & 0.17 & 1.33 & 2.12 & 2.38 & 0.01 &  0 &  0 \\ 
10&  0.00 & 0.70 & 0.69 & 1.41 & 3.21 & 6.76 &  0 &  8.7 \\ 
12&  0.00 & {\it 0.00} & {\it 0.52} & 1.69 & {\bf 3.78} & 32.64 & 33.3 & 31.0  \\   \hline  \hline
\end{tabular}

\newpage

{TABLE II. The dimension of two-particle coefficients of fractional parentage (cfp's) in
the case of single-$j=\frac{15}{2}$ with 5 fermions.
The total dimension of two-body cfp's for different $I$ and
that of the shell model space  are given in the two columns
``cfp's" and ``SM", respectively.
Note that the  dimension  distribution  of two-body cfp's for 
fermions in a larger single-$j$ shell or a   many-$j$ shell  is 
similar. The ${\cal P}(I)$'s  
  in the last column (in $\%$)  are obtained
by diagonalizing   a TBRE hamiltonian 1000 runs.
Bold font is used for the  $d_I^J/D_I$'s which
give  the largest, and italic font is used
for those which are very comparable. }

\begin{tabular}{cccccccccccc} \hline  \hline
$2I$   &  $G_0$ &  $G_2$ &  $G_4$ &  $G_6$ &  $G_8$ & $G_{10}$ & $G_{12}$  & $G_{14}$  & cfp's  &  SM & ${\cal P}(I)$ \\  \hline 
1 & 0  & {\bf 2} & {\it 3} & 4 & {\bf 5}  &  {\it 4} & 3 & 3 & 24  & 2 & 18.4 \\
3 & {\bf 1}  & 3 & {\it 6} & {\bf 9} & 9  &  {\it 8} & 7 & 5 & 48  & 4 & 12.8 \\
5 & 1  & {\it 5} & {\it 9} &12 &13  & {\it 12} &10 & 8 & 70  & 6 & 12.6 \\
7 & 1  & 7 &{\it 12} &15 &17  & {\it 16} &13 &11 & 92  & 8 & 3.6 \\
9 & 2  & 8 &{\it 14} &18 &19  & {\it 19} &17 &13 &110  & 9 & 1.5 \\
11& 2  &10 &{\it 16} &20 &22  & {\it 22} &20 &16 &128  &11 & 1.4 \\
13& 2  &11 &{\it 17} &21 &24  & {\it 24} &22 &19 &140  &11 & 0.1 \\
15& 3  &11 &17 &23 &25  & {\it 26} &25 &22 &152  &13 & 0.1 \\
17& 2  &11 &18 &23 &26  & {\it 27} &26 &23 &156  &12 & 0\\
19& 2  &11 &18 &23 &27  & {\it 28} &27 &25 &161  &13 & 0\\
21& 2  & 9 &17 &23 &26  & {\it 27} &28 &27 &159  &12 & 0\\
23& 2  & 9 &16 &22 &26  & {\it 28} &28 &27 &158  &12 & 0\\
25& 1  & 8 &14 &20 &25  & {\it 27} &28 &28 &151  &11 & 0\\
27& 2  & 7 &13 &19 &23  & {\it 26} &28 &28 &146  &11 & 0\\
29& 1  & 6 &11 &16 &21  & {\it 25} &27 &28 &135  & 9 & 0\\
31& 1  & 6 &10 &14 &19  & {\it 23} &26 &28 &127  & 9 & 0\\
33& 1  & 4 & 8 &12 &16  & {\it 21} &25 &27 &114  & 7 & 0\\
35& 1  & 4 & 7 &10 &14  & {\it 19} &23 &26 &104  & 7 & 0.1\\
37& 0  & 3 & 5 & 8 &12  & {\it 16} &21 &25 & 90  & 5 & 0\\
39& 1  & 2 & 4 & 7 &10  & {\it 14} &19 &23 & 80  & 5 & 0.4\\
41& 0  & 2 & 3 & 5 & 8  & {\it 12} &16 &21 & 66  & 3 & 0\\
43& 0  & 2 & 2 & 4 & 7  & {\it 10} &14 &19 & 57  & 3 & 0\\
45& 0  & 0 & 1 & 3 & 5  & {\it 8} &12 &16 & 45  & 2 & 2.0\\
47& 0  & 0 & 1 & 2 & 4  & {\it 7} &10 &14 & 38  & 2 & 2.3\\
49& 0  & 0 & 0 & 1 & 3  & {\it 5} & 8 &12 & 29  & 1 & 5.5\\
51& 0  & 0 & 0 & 1 & 2  & {\it 4} & {\bf 7} &10 & 24  & 1 & 10.3\\
55& 0  & 0 & 0 & 0 & 1  & 2 & 4 & {\bf 7} &  14  & 1 & 28.9\\  \hline  \hline
\end{tabular}

\newpage

{\bf Figure captions}: 

\vspace{0.25in}

FIG.~1 ~~ Typical behaviors of the ${\cal P}(I)$'s:    
a)~ single-$j$ ($j=\frac{15}{2}$) with 4 fermions,
b)~ two-$j$ shell ($2j_1,2j_2)=(11,7)$ with 6 fermions,
c)~ 6-$sdg$ bosons,  
d)~ single-$j$ ($j=\frac{9}{2}$) with 5 fermions,

\vspace{0.25in}

FIG. ~ 2 ~~ Typical distribution of  two-particle cfp's  and
$\alpha_{I \beta \beta'}^J$ of
 4 fermions in a $j=31/2$ shell and a medium angular momentum $I$ (=20).
  It is seen that the distributions are
 close to Poissonian. The   distributions of other
 $I$ are similar unless $I\sim I_{min}$, or $I_{max}$, where
 the numbers of the two-particle cfp's and $\alpha_{I \beta \beta'}^J$
 are small.  

\vspace{0.25in}

FIG.~3 ~~ Typical behavior of  $\langle \bar{E}_I\rangle_{min}$ vs. $I(I+1)$.
 The $\langle \bar{E}_I\rangle_{min}$'s are  obtained
by averaging over all $\bar{E}_I$  given by 
diagonalizing a TBRE hamiltonian,  with a requirement that
 $I \sim I_{min}$  g.s. 
a). 20-$d$ bosons; b) 10-$sd$ bosons;
c) single-$j$ shell with $j$=17 and $n=4$,
and d)  a two-$j$ shell with $j_1=5/2, j_2 =7/2$, and  $n=$4.  
The reversed cases, i.e., the $I\sim I_{max}$  g.s.,
are very similar to the cases of $I\sim I_{min}$  g.s..

\vspace{0.25in}

FIG.~4 ~~Coefficients $C$ (in  $\langle \bar{E}_I\rangle_{min}$$=C I (I+1)$)
of fermions in  different shells.
It is seen that $C$ is not sensitive to particle number $n$, but rather
sensitive to $j$ values of the shell. Refer text for details.

\vspace{0.25in}

FIG.~5 ~~Correlation between $\sqrt{\cal J}$ and $j=\sqrt{ \sum_i j^2_i}$.
It is indicated that ${\cal J}$ is approximately proportional to
 ${ \sum_i j^2_i}$ for $d$- and $sd$-bosons, and fermions
 in a single-$j$ shell, with   ${\cal J}$ of many-$j$ shells  and $sdg$-boson
shifted  very slightly to the right.

\vspace{0.25in}

FIG.~6 ~~ A scenario of $\bar{E}_{I} \sim I(I+1)$.
 for 4 fermions in a single-$j$ ($j=15/2$) shell.
The dot straight lines are used to guide the eyes.
 Only even $I$ are plotted, because the predicted $\bar{E}_{I+1}$
 ($I$ is even) is the same as  the predicted $\bar{E}_{I}$.
Solid line is obtained by a linear fit.
 a) Solid squares  are obtained by
using predicted $\bar{\alpha}_I^J$
($=d_I^J/D_I \times \frac{1}{2} n (n-1)$) and
 taking $G_J$'s to be TBRE. 
 b) Solid squares are obtained by using the same $\bar{\alpha}_I^J$
with a simple assumption of $G_J$. Refer to the text for details.

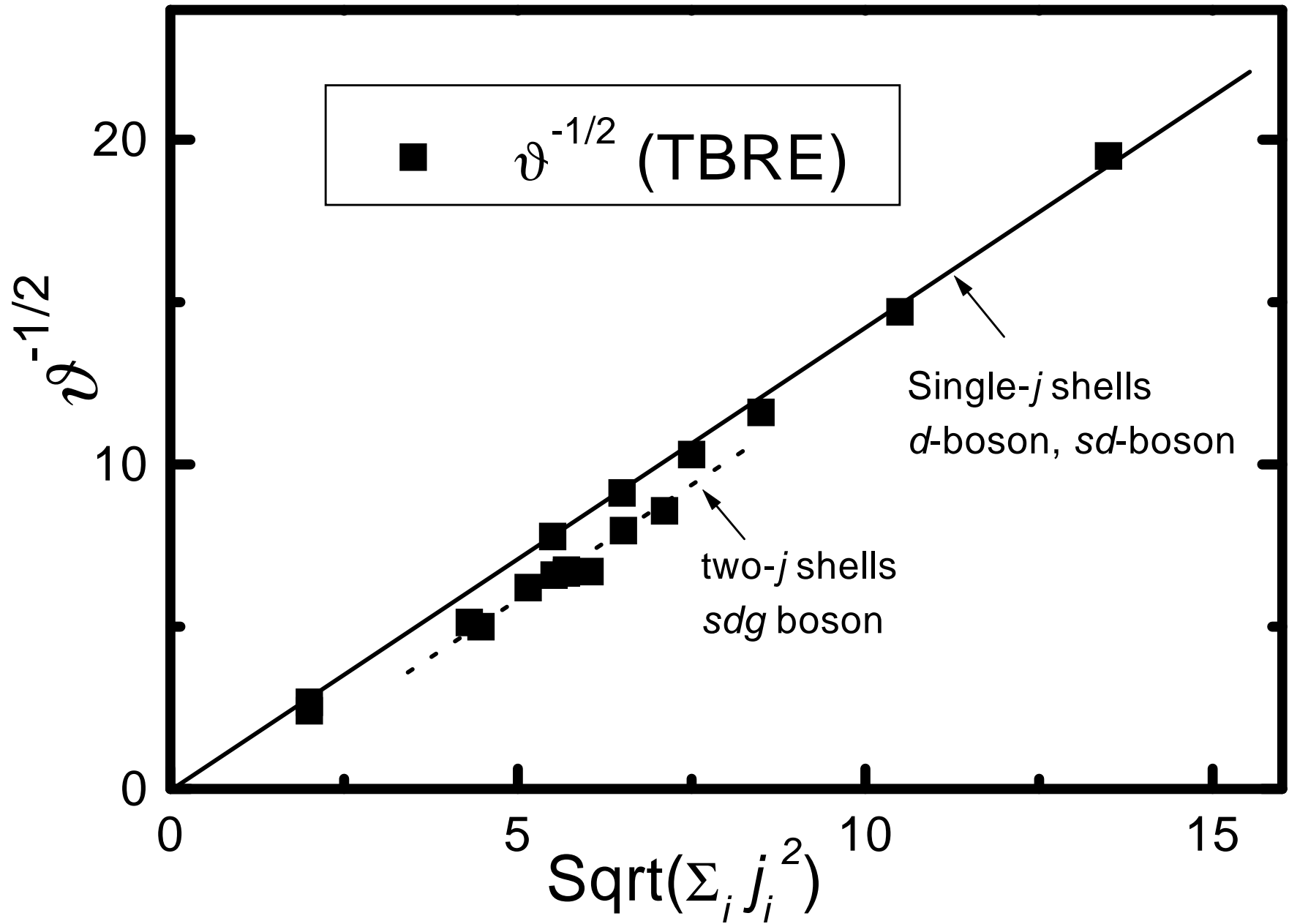
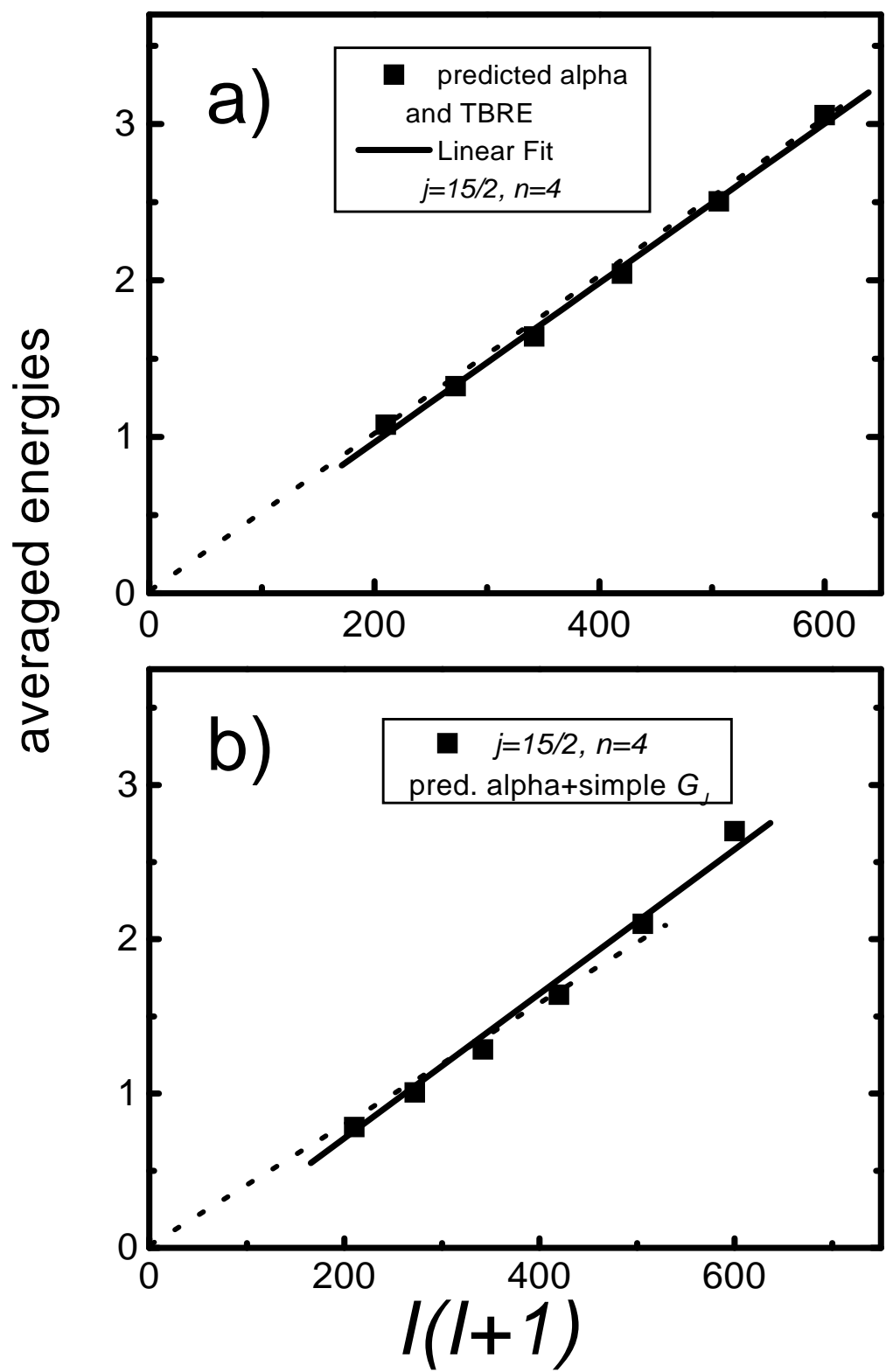
\end{document}